\documentclass[multi]{cambridge7A}
\usepackage[UKenglish]{babel}
\usepackage[longnamesfirst,sectionbib]{natbib}
\usepackage{chapterbib}
\usepackage{amssymb,amsmath}
\usepackage{amsthm}
\usepackage{graphicx}
\usepackage{enumerate,url}
\usepackage[utf8]{inputenc}

\newcommand{\be}{\begin{equation}}
\newcommand{\bel}[1]{\begin{equation}\label{#1}}
\newcommand{\ee}{\end{equation}}
\newcommand{\BC}{\begin{center}}
\newcommand{\EC}{\end{center}}
\newcommand{\BI}{\begin{itemize}}
\newcommand{\EI}{\end{itemize}}
\newcommand{\BE}{\begin{enumerate}}
\newcommand{\EE}{\end{enumerate}}
\newcommand{\I}{\item}

\newcommand{\R}{\mathcal{R}}

\newcommand{\cv}{\mbox{\boldmath$c$}}

\setcitestyle{authoryear,round,semicolon}

\allowdisplaybreaks[1]

\providecommand*\Index[1]{#1\index{#1}}
\providecommand*\undex[1]{} 
\providecommand*\Undex[1]{#1} 

\begin{document}
\alphafootnotes
\author[Peter J. Green]
{Peter J. Green\footnotemark}
\chapter[Colouring and breaking sticks]
{Colouring and breaking sticks: random distributions and heterogeneous clustering}
\footnotetext[1]{School of Mathematics, University of Bristol, Bristol BS8 1TW;
  P.J.Green@bristol.ac.uk}
\arabicfootnotes
\contributor{Peter J. Green \affiliation{University of Bristol}}
\renewcommand\thesection{\arabic{section}}
\numberwithin{equation}{section}
\renewcommand\theequation{\thesection.\arabic{equation}}
\numberwithin{figure}{section}
\renewcommand\thefigure{\thesection.\arabic{figure}}

\begin{abstract}
We begin by reviewing some probabilistic results about the Dirichlet Process and its close relatives,
focussing on their implications for statistical modelling and analysis. We then introduce
a class of simple mixture models in which clusters are of different `colours', with statistical characteristics that are constant within colours, but different between colours. Thus cluster identities are
exchangeable only within colours. The basic form of our model is a variant on the
familiar Dirichlet process, and we find that much of the standard modelling
and computational machinery associated with the Dirichlet process may be readily adapted to our generalisation.
The methodology is illustrated with an application to the partially-parametric clustering of gene expression profiles.
\end{abstract}

\subparagraph{Keywords}
Bayesian nonparametrics,
gene expression profiles,
hierarchical models,
loss functions,
MCMC samplers,
optimal clustering,
partition models,
P\'olya urn,
stick breaking

\subparagraph{AMS subject classification (MSC2010)}60G09, 62F15, 62G99, 62H30, 62M99

\section{Introduction}
The purpose of this note is four-fold: to remind some Bayesian\index{Bayes, T.!Bayesian statistics} nonparametricians gently that closer study of some
probabilistic literature might be rewarded, to encourage
probabilists\index{probability} to think that there are statistical modelling
problems worth of their attention, to point out to all another important
connection between the work of John Kingman and modern statistical methodology
(the role of the coalescent\index{Kingman, J. F. C.!Kingman coalescent} in
population genetics\index{mathematical genetics} approaches to \Index{statistical genomics} being the most important example;
see papers by Donnelly\index{Donnelly, P. [Donnelly, P. J.]}, Ewens\index{Ewens, W. J.} and
Griffiths\index{Griffiths, R. C.} in this volume), and finally to introduce a
modest generalisation of the Dirichlet
process\index{Dirichlet, J. P. G. L.!Dirichlet process (DP)|(}.

The most satisfying basis for statistical clustering of items of data is a probabilistic model, which usually takes the form of a mixture model\index{mixture model|(}, broadly interpreted. In most cases, the statistical characteristics of each cluster or mixture component are the same, so that cluster identities are \emph{a priori} exchangeable. In Section \ref{sec:cdp} we will introduce
a class of simple mixture models in which clusters are of different categories, or colours as we shall call them,  with statistical characteristics that are constant within colours, but different between colours. Thus cluster identities are
exchangeable\index{mixture model!exchangeability} only within colours.

\section{Mixture models and the Dirichlet process}
\label{sec:mixdp}

Many statistical models have the following character. Data $\{Y_i\}$ are
available on $n$ units that we shall call \emph{items}, indexed $i=1$, 2,
\ldots, $n$. There may be item-specific covariates\index{covariate}, and other
information, and the distribution of each $Y_i$ is determined by an unknown
parameter $\phi_i\in\Omega$, where we will take $\Omega$ here to be a subset
of a Euclidean space. Apart from the covariates, the items are considered to
be exchangeable, so we assume the $\{Y_i\}$ are conditionally
independent\index{conditional independence} given $\{\phi_i\}$, and model the $\{\phi_i\}$ as exchangeable\index{exchangeability|(} random variables. Omitting covariates for simplicity, we write $Y_i|\phi_i \sim f(\cdot|\phi_i)$.

It is natural to take $\{\phi_i\}$ to be independent and identically
distributed random variables, with common distribution $G$, where $G$ itself
is unknown, and treated as random. We might be led to this assumption whether
we are thinking of a de Finetti-style\index{Finetti, B. de} representation
theorem (\citet{DeFinetti31,DeFinetti37}; see also
\citet{KingmanExch}\index{Kingman, J. F. C.},
\citet{Kallenberg})\index{Kallenberg, O.}, or by following \Index{hierarchical
modelling} principles \citep{GelmanCSR,GreenHSSS}\index{Gelman, A.}\index{Hjort, N. L.}\index{Richardson, S.}, 
Thus, unconditionally, $Y_i|G \sim \int f(\cdot|\phi) G(d\phi)$, independently given $G$.

This kind of formulation enables us to borrow strength across the units in inference about unknown parameters, with the aim of controlling the \Index{degrees of freedom}, capturing the idea that while the $\{\phi_i\}$ may be different from item to item, we nevertheless understand that, through exchangeability\index{exchangeability|)}, knowing the value of one of them would tell us something about the others.

There are still several options. One is to follow a standard parametric formulation, and to assume a specific parametric form for $G$, with parameters, or rather `hyperparameters'\index{hyperparameter}, in turn given a hyperprior distribution. However, many would argue that in most practical contexts, we would have little information to build such a model for $G$, which represents variation in the population of possible items of the parameter $\phi$ that determines the distribution of the data $Y$. 

Thus we would be led to consider more flexible models, and one of several approaches might occur to us:
\BI
\I a nonparametric approach, modelling uncertainty about $G$ without making parametric assumptions;
\I a mixture model\index{mixture model|)} representation for $G$;
\I a partition model\index{partition!model|(}, where the $\{\phi_i\}$ are grouped together, in a way determined \emph{a posteriori} by the data.
\EI
One of the things we will find, below, is that taking natural choices in each of these approaches can lead to closely related formulations in the end, so long as both modelling and inference depend solely on the $\{\phi_i\}$. These connections, not novel but not entirely well-known either, shed some light on the nature and implications of the different modelling approaches.

\subsection{Ferguson definition of the Dirichlet process}
\label{sec:ferguson}
Much Bayesian nonparametric distributional modelling\index{Bayes, T.!Bayesian nonparametric modelling|(} \citep{WalkerDLS} begins with the Dirichlet process
\citep{Ferguson}\index{Ferguson, T. S.}. Building on earlier work by
Dubins\index{Dubins, L.}, Freedman\index{Freedman, D. A.} and
Fabius\index{Fabius, J.}, Ferguson intended this model to provide a nonparametric prior model for $G$ with a large support, yet one
remaining capable of tractable prior-to-posterior analysis.

Given a probability distribution $G_0$ on an arbitrary
measure space $\Omega$, and a positive real $\theta$, we say
the random distribution $G$ on $\Omega$ follows a Dirichlet process,
$$
G\sim DP(\theta, G_0),
$$
if for all partitions $\Omega =\bigcup_{j=1}^m B_j$ ($B_j\cap B_k=\emptyset$
if $j\neq k$), and for all $m$,
$$
(G(B_1),\ldots,G(B_m))\sim \mbox{Dirichlet}
(\theta G_0(B_1),\ldots,\theta G_0(B_m)),
$$
where Dirichlet$(\alpha_1,\alpha_2,\ldots,\alpha_m)$\index{Dirichlet, J. P. G. L.!Dirichlet distribution|(} denotes the distribution
on the $m$-dim\-ensional
simplex with density at $(x_1,x_2,\ldots,x_m)$ proportional to
$\prod_{j=1}^m\allowbreak x_j^{\alpha_j-1}$.

The \Index{base measure} $G_0$ gives the \emph{expectation} of $G$:
$$
E(G(B)) = G_0(B).
$$

Even if $G_0$ is continuous, $G$ is a.s. discrete
\citep{KingmanPac,Ferguson,Blackwell,KingmanRSS}\index{Kingman, J. F. C.|(}\index{Blackwell, D.},
so i.i.d. draws $\{\phi_i,i=1,2,\ldots,n\}$ from $G$ exhibit ties. 
The parameter $\theta$ measures (inverse) \emph{concentration}\index{concentration}: given
i.i.d. draws $\{\phi_i,i=1,2,\ldots,n\}$ from $G$,
\BI
\I as $\theta\to 0$, all $\phi_i$ are equal, a single draw from $G_0$;
\I as $\theta\to\infty$, the $\phi_i$ are drawn i.i.d. from $G_0$.
\EI

\subsection{The stick-breaking construction}
\label{sec:stick}
A draw $G$ from a Dirichlet process is a discrete distribution on $\Omega$, so an alternative way to define the Dirichlet process would be via a construction of such a random distribution, through specification of the joint distribution of the locations of the atoms, and their probabilities. Such a construction
was given by \citet{Ferguson}: in this, the locations are i.i.d. draws from $G_0$, with probabilities forming a decreasing sequence constructed from increments of a \Index{gamma process}.

This is not the explicit construction that is most commonly used today, which
is that known in the Bayesian
nonparametric community
as Sethuraman's\index{Sethuraman, J.!Sethuraman stick-breaking model} stick-breaking
model \citep{SethuramanT,Sethuraman}\index{Tiwari, R. C.}. This leads to this algorithm for generating the $\{\phi_i\}$:
\BE[1.]
\I draw $\phi^\star_j\sim G_0$, i.i.d., $j=1$, 2, \ldots;
\I draw $V_j\sim \mbox{Beta}(1,\theta)$, i.i.d., $j=1$, 2, \ldots;
\I define $G$ to be the discrete distribution putting
probability $(1-V_1)(1-V_2)\ldots(1-V_{j-1})V_j$
on $\phi^\star_j$;
\I draw $\phi_i$ i.i.d.\ from $G$, $i=1$, 2, \ldots, $n$.
\EE

This construction can be found considerably earlier in the probability
literature, especially in connection with models for
\index{sampling!species sampling}species sampling. The earliest reference
seems to be in
\citet{McCloskey}\index{McCloskey, J. W.}; for more readily accessible
sources, see \citet{PatilT}\index{Patil, C. P.}\index{Taillie, C.} and
\citet{DonnellyJ}\index{Donnelly, P. [Donnelly, P. J.]}\index{Joyce, P.}, where it is
described in the context of size-biased sampling\index{sampling!size-biased sampling}
and the GEM (Generalised Engen--McCloskey) distributions\index{GEM distribution}. See also Section \ref{sec:general} below.

\subsection{Limits of finite mixtures}
\label{sec:finitemixture}
A more direct, classical approach to modelling the distribution of $Y$ in a flexible way would be to use a finite \Index{mixture model}. Suppose that $Y_i$ are i.i.d. with density $\sum_j w_j f_0(\cdot|\phi^\star_j)$ for a prescribed parametric density family $f_0(\cdot|\phi)$, and consider a Bayesian formulation\index{Bayes, T.!Bayesian nonparametric modelling|)} with priors on the component weights $\{w_j\}$ and the component-specific parameters $\{\phi_j^\star\}$. The simplest formulation (e.g. 
\citet{RichardsonG})\index{Richardson, S.} uses a Dirichlet prior\index{Dirichlet, J. P. G. L.!Dirichlet distribution|)} on the weights, and takes the $\{\phi_j^\star\}$ to be i.i.d. \emph{a priori}, but with arbitrary distribution, so in algorithmic form:
\BE[1.]
\I draw $(w_1,w_2,\ldots,w_k)\sim\mbox{Dirichlet}(\delta,\ldots,\delta)$;
\I draw $c_i\in\{1,2,\ldots,k\}$ with $P\{c_i=j\}=w_j$, i.i.d., $i=1$, \ldots,
$n$;
\I draw $\phi^\star_j\sim G_0$, i.i.d., $j=1$, \ldots, $k$;
\I set $\phi_i=\phi^\star_{c_i}$.
\EE

It is well known that if we take the limit 
$k\to\infty$, $\delta\to 0$ such that $k\delta\to \theta$, then the joint distribution of the $\{\phi_i\}$ is the same as that obtained via the Dirichlet process formulation in the previous subsections (see for example \citet{GreenR}). This result is actually a corollary of a much stronger statement due to \citet{KingmanRSS}, about the convergence of discrete probability measures.
For more recent results in this direction see
\citet{Muliere}\index{Muliere, P.}\index{Secchi, P.} and
\citet{IshwaranZ}\index{Ishwaran, H.}\index{Zarepour, M.}.

We are still using the formulation $Y_i|G \sim \int f(\cdot|\phi) G(d\phi)$, independently given $G$, but note that $G$ is invisible in this view; it has implicitly been integrated out.

\subsection{Partition distribution}
\label{sec:partition}
Suppose that, as above, $G$ is drawn from $DP(\theta,G_0)$, and then $\{\phi_i:i=1,2,\ldots,n\}$
drawn i.i.d. from $G$. We can exploit the conjugacy\index{conjugacy} of the
Dirichlet with respect to multinomial sampling\index{sampling!multinomial sampling} to integrate out $G$.
For a fixed partition $\{B_j\}_{j=1}^m$ of $\Omega$, and integers $c_i\in\{1,2,\ldots,m\}$,
we can write
$$
P\{\phi_i\in B_{c_i},i=1,2,\ldots,n\} = \frac{\Gamma(\theta)}{\Gamma(\theta+n)}
\prod_{j=1}^m \frac{\Gamma(\theta G_0(B_j)+n_j)}{\Gamma(\theta G_0(B_j))},
$$
where $n_j=\#\{i:c_i=j\}$. The $j$th factor in the product above is 1 if $n_j=0$, and otherwise
$\theta G_0(B_j)(\theta G_0(B_j)+1)(\theta G_0(B_j)+2)\ldots(\theta G_0(B_j)+n_j-1)$, so we find that if the partition becomes increasingly refined, and $G_0$ is non-atomic, then the joint distribution of the
$\{\phi_i\}$ can equivalently be described by 
\begin{enumerate}[1.]
\I partitioning $\{1,2,\ldots,n\}=\bigcup_{j=1}^d C_j$
at random, so that
\bel{dppart}
p(C_1,C_2,\ldots,C_d) = \frac{\Gamma(\theta)}{\Gamma(\theta+n)}
\theta^d \prod_{j=1}^d (n_j-1)!
\ee
where $n_j=\#C_j$;
\I drawing $\phi^\star_j\sim G_0$, i.i.d., $j=1$, \ldots, $d$, and then
\I setting $\phi_i=\phi^\star_j$ if $i\in C_j$.
\end{enumerate}

Note that the partition model\index{random partition|(} (\ref{dppart}) shows
extreme preference for unequal cluster sizes\index{cluster(ing)!cluster size}. 
If we let $a_r=\#\{j:n_j=r\}$, then the joint distribution of $(a_1,a_2,\ldots)$ is  
\bel{ewens1}
\frac{n!}{n_1!n_2!\cdots n_d!}\times \frac{1}{\prod_r a_r!} \times p(C_1,C_2,\ldots,C_d).
\ee
This is equation (A3) of
\citet{Ewens}\index{Ewens, W. J.!Ewens sampling formula (ESF)}, derived in a
context where $n_j$ is the number of genes\index{gene} in a sample of the
$j$th allelic type, in \Index{sampling} from a
\index{selection!selectively neutral}selectively neutral population process. 
The first factor in (\ref{ewens1}) is the multinomial coefficient accounting for the number of ways the $n$ items can be allocated to clusters of the required sizes, and the second factor accounts for the different sets of
$\{n_1,n_2\ldots,n_d\}$ leading to the same $(a_1,a_2,\ldots)$. Multiplying all this together, a little manipulation leads to the familiar Ewens sampling formula:
\bel{ewens}
p(a_1,a_2,\ldots) = \frac{n! \Gamma(\theta)}{\Gamma(\theta+n)} \prod_r \frac{\theta^{a_r}}{r^{a_r}a_r!}.
\ee
See also \citet{KingmanPP}\index{Kingman, J. F. C.|)}, page 97.

This representation of the partition structure implied by the Dirichlet process 
was derived by \citet{Antoniak}\index{Antoniak, C. E.}, in the form (\ref{ewens}). He noted that a consequence of this representation is that the joint distribution of the $\{\phi_i\}$ given $d$
is independent for $\theta$; thus given observed $\{\phi_i\}$, 
$d$ is sufficient\index{sufficient statistic} for $\theta$. A similar observation was also made by \citet{Ewens}
in the genetics\index{mathematical genetics} context of his work.

Note that as in the previous section, $G$ has been integrated out, and so is invisible in this view of the Dirichlet process model.

\subsection{Reprise}
Whichever of the points of view is taken, items are clustered, according to a tractable distribution 
parametrised by $\theta>0$, and for each cluster the cluster-specific
parameter $\phi$ is an independent draw from $G_0$. Much statistical
methodology built on the Dirichlet-process model uses only this joint distribution of the $\{\phi_i\}$, and so should hardly be called `nonparametric'. Of course, even though $G$ itself is invisible in two of the derivations above, the Dirichlet-process model does support inference about $G$, but this is seldom exploited in applications.

\subsection{Multiple notations for partitions}

In what follows, we will need to make use of different notations for the random partition\index{random partition|)} induced by the Dirichlet-process model, or its relatives. We will variously use
\BI 
\I $\cv$ is a partition of $\{1,2,\ldots,n\}$;
\I clusters of partition are $C_1$, $C_2$, \ldots, $C_d$ ($d$ is the \emph{degree}\index{partition!degree}
of the partition): $\bigcup_{j=1}^d C_j=\{1,2,\ldots,n\}$,
$C_j \cap C_{j'}=\emptyset$ if $j\neq j'$;
\I $c$ is the \Undex{allocation vector}: $c_i=j$ if and only if $i\in C_j$.
\EI

Note that the first of these makes no use of the (arbitrary) labelling of the clusters used in the second and third. We have to take care with multiplicities, and the distinction between (labelled) allocations and (unlabelled) partitions.

\section{Applications and generalisations}

\subsection{Some applications of the Dirichlet process in Bayesian nonparametrics}

Lack of space precludes a thorough discussion of the huge statistical
methodology literature exploiting the Dirichlet process in Bayesian
nonparametric procedures, so we will only review a few highlights\index{Bayes, T.!Bayesian nonparametric modelling|(}.

\citet{Lo}\index{LoAY@Lo, A. Y.} proposed
\index{density estimation|(}density estimation procedures
devised by mixing a user-defined kernel function $K(y,u)$ with respect to a
Dirichlet process; thus i.i.d. data $\{Y_i\}$ are assumed distributed as $\int
K(\cdot,u) G(du)$ with $G$ drawn from a Dirichlet process. This is now known
as the \index{Dirichlet, J. P. G. L.!Dirichlet process mixture model}Dirichlet process mixture model (a better terminology than the formerly-used `mixture of Dirichlet processes'). The formulation is identical to that we started with in Section \ref{sec:mixdp}, but for the implicit assumption that $y$ and $u$ lie in the same space, and that the kernel $K(\cdot,u)$ is a \Index{unimodal density} located near $u$.

In the 1990s there was a notable flourishing of applied Bayesian
nonparametrics, stimulated by interest in the Dirichlet process, and the rapid
increase in computational power available to researchers, allowing almost
routine use of the P\'olya urn\index{Polya, G.@P\'olya, G.!P\'olya urn} sampler approach (see Section \ref{sec:polya}) to posterior computation. 
For example, \citet{Escobar}\index{Escobar, M. D.} re-visited the Normal Means
problem\undex{normal means problem},
\citet{WestME}\index{West, M.}\index{Mueller, P.@M\"uller, P.} discussed regression\index{regression}
and density estimation, and \citet{EscobarW} further developed Bayesian
\index{density estimation|)}density
estimation\index{Bayes, T.!Bayesian density estimation}.
\citet{MuellerEW}\index{Erkanli, A.} ingeniously exploited multivariate
density estimation using Dirichlet process mixtures to perform Bayesian curve
fitting\index{Bayes, T.!Bayesian curve fitting} of one margin on the others.

\subsection{Example: clustered linear models for gene expression profiles}
\label{sec:gene}



Let us consider a substantial and more specific application in some detail, to
motivate the Dirichlet process (DP) set-up as a natural elaboration of a
standard parametric Bayesian hierarchical
model\index{Bayes, T.!Bayesian hierarchical model} approach\index{Bayes, T.!Bayesian nonparametric modelling|)}.

A remarkable aspect of modern \Index{microbiology} has been the dramatic development
of novel high-throughput assays\index{assay}, capable of delivering very high
dimensional quantitative
\index{data!availability}data on the genetic characteristics of organisms from
biological samples. One such technology is the measurement of gene
expression\index{gene!expression|(} using Affymetrix gene
chips\index{Affymetrix gene chip@\emph{Affymetrix} gene chip}. In
\citet{LauG}\index{LauJW@Lau, J. W.|(}, we work with possibly replicated gene expression measures. The data are
$\{Y_{isr}\}$, indexed by
\BI
\I genes $i=1$, 2, \ldots, $n$,
\I conditions $s=1$, 2, \ldots, $S$, and
\I replicates $r=1$, 2, \ldots, $R_s$.
\EI
Typically $R_s$ is very small,
$S$ is much smaller than $n$, and the `conditions' represent
different subjects, different treatments, or different experimental
settings. 

We suppose there is a $k$-dimensional ($k\leq S$) \Index{covariate} vector
$x_s$ describing each condition, and model parametric dependence
of $Y$ on $x$; the focus of interest is on the pattern of variation in these gene-specific parameters across the
assayed genes.

Although other variants are easily envisaged, we suppose here that
$$
Y_{isr}\sim N(x_s'\beta_i,\tau_i^{-1}),\quad\mbox{independently}.
$$
Here $\phi_i=(\beta_i,\tau_i)\in\R^{k+1}$ is a gene-specific parameter vector characterising the dependence of gene expression\index{gene!expression|)} on the condition-specific covariates. \emph{A priori}, the genes can be considered exchangeable, and a standard hierarchical formulation would model the $\{\phi_i\}$ as i.i.d. draws from a parametric prior distribution $G$, say, whose (hyper)parameters have unknown values. This set-up allows borrowing of strength across genes in the interest of stability and efficiency of inference. 

The natural nonparametric counterpart to this would be to suppose instead that $G$, the distribution describing variation of $\phi$ across the population of genes, does not have prescribed parametric form, but is modelled as a random distribution from a `nonparametric' prior such as the Dirichlet process, specifically 
$$
G\sim DP(\theta, G_0).
$$
A consequence of this assumption, as we have seen, is that $G$ is atomic, so that the genes will be clustered together into groups sharing a common value of $\phi$. \emph{A posteriori} we obtain a probabilistic clustering of the gene expression profiles. 

\citet{LauG}\index{LauJW@Lau, J. W.|)} take a standard
normal--inverse Gamma model\index{normal--inverse gamma model}, so that $\phi=(\beta,\tau)\sim G_0$ means
$$
\tau\sim\Gamma(a_0,b_0) \quad\mbox{and}\quad \beta|\tau\sim \mbox{N}_k(m_0,(\tau t_0)^{-1}I).
$$

This is a conjugate\index{conjugacy} set-up, so that $(\beta,\tau)$ can be integrated out
\emph{in each cluster}.
This leads easily to explicit within-cluster parameter posteriors:
\begin{gather*}
\tau^\star_j|Y \sim \Gamma(a_j,b_j),
\\
\beta^\star_j|\tau^\star_j,Y \sim \mbox{N}_k(m_j,(\tau^\star_j t_j)^{-1}),
\end{gather*}
where
\begin{gather*}
a_j = a_0+1/2\#\{isr:c_i=j\},
\\
b_j = b_0+1/2(Y_{C_j}-X_{C_j}m_0)'(X_{C_j}t_0^{-1}X_{C_j}')^{-1}(Y_{C_j}-X_{C_j}m_0),
\\
m_j = (X'_{C_j}X_{C_j}+t_0I)^{-1}(X'_{C_j}Y_{C_j}+t_0m_0),
\\
t_j = X'_{C_j}X_{C_j}+t_0I.
\end{gather*}
The marginal likelihoods\index{likelihood!marginal likelihood} $p(Y_{C_j})$ are
multivariate $t$ distributions\index{multivariate $t$ distribution}.

We continue this example later, in Sections \ref{sec:clusreg} and \ref{sec:timecourse}.

\subsection{Generalisations of the Dirichlet process, and related models}
\label{sec:general}

Viewed as a nonparametric model or as a basis for probabilistic clustering\index{cluster(ing)}, the Dirichlet process is simple but inflexible---a single real parameter $\theta$ controls both variation and \Index{concentration}, for example. And although the space $\Omega$ where the \Index{base measure} $G_0$ lies and in which $\phi$ lives can be rather general, it is essentially a model for `univariate' variation and unable to handle in a flexible way, for example, \Index{time-series} data.

Driven both by such considerations of statistical\index{statistics} modelling
\citep{WalkerDLS}\index{Walker, S. G.|(}\index{Damien, P.|(}\index{Laud, P. W.|(}\index{Smith, A. F. M.|(}, or curious pursuit of more general mathematical results, the Dirichlet process has proved a fertile starting point for numerous generalisations, and we touch on just a few of these here.

\paragraph{The Poisson--Dirichlet distribution and its two-parameter generalisation.}
\citet{KingmanRSS}\index{Kingman, J. F. C.} observed and exploited the fact
that the limiting behaviour of random discrete
distributions\index{random discrete distribution} could become non-trivial and accessible through
permutation of the components to be in ranked (decreasing) order. The limit
law is the Poisson--Dirichlet distribution\index{Poisson, S. D.!PoissonDirichlet distribution@Poisson--Dirichlet distribution}, implicitly defined and later described \citep[page 98]{KingmanPP} as `rather less than user-friendly'. 

\citet{DonnellyJ}\index{Donnelly, P. [Donnelly, P. J.]}\index{Joyce, P.} elucidated the role
of both \Index{ranking} and size-biased sampling\index{sampling!size-biased sampling}
in establishing limit laws for random distributions; see also
\citet{Holst}\index{Holst, L.} and \citet[page 107]{ArratiaBT}\index{Arratia, R.}\index{Barbour, A. D.}\index{Tavar\'e, S.}. 
The two-parameter generalisation of the Poisson--Dirichlet model was
discovered by Pitman and co-workers, see for example
\citet{PitmanY}\index{Pitman, J. [Pitman, J. W.]}\index{Yor, M.}. This has been a rich topic for probabilistic study to the present day; see chapters by 
Gnedin\index{Gnedin, A. V.}, Haulk\index{Haulk, C.} and Pitman, and by
Aldous\index{Aldous, D. J.} in this volume. The simplest view to take of the
two-parameter Poisson--Dirichlet model\index{two-parameter Poisson--Dirichlet process} PD$(\alpha,\theta)$ is to go back to
stick-breaking\index{stick breaking} (Section \ref{sec:stick}) and replace the
Beta$(1,\theta)$ distribution for the variables $V_j$ there by
Beta$(1-\alpha,\theta+j\alpha)$\index{beta distribution|(}. 

\citet{IshwaranJ01}\index{Ishwaran, H.}\index{James, L. F.} have considered
Bayesian statistical applications of stick-breaking priors defined in this
way, and implementation of Gibbs sampling\index{Gibbs, J. W.!Gibbs sampler} for computing posterior distributions.

\paragraph{Dirichlet process relations in structured dependent models.}
Motivated by the need to build statistical models for \Index{structured data} of various kinds, there has been a huge effort in generalising Dirichlet process models for such situations---indeed, there is now an `$x$DP' for nearly every letter of the alphabet. 

This has become a rich and sometimes confusing area; perhaps the most important current models are 
Dependent Dirichlet processes\index{Dirichlet, J. P. G. L.!Dirichlet process generalisations}
\citep{MacEachernASA,MacEachernKG}\index{MacEachern, S. N.}\index{Kottas, A.}\index{Gelfand, A.},
Order-based dependent Dirichlet processes
\citep{GriffinS}\index{Griffin, J. E.}\index{Steel, M. F. J.},
Hierarchical Dirichlet processes
\citep{TehJBB}\index{Teh, Y. W.}\index{Jordan, M. I.}\index{Beal, M. J.}\index{Blei, D. M.}, and
Kernel stick breaking processes\index{kernel stick-breaking process}
\citep{DunsonP}\index{Dunson, D. B.}\index{Park, J.-H.}. Many of the models are based on stick-breaking representations, but in which the atoms and/or the weights for the representations of different components of the process are
made dependent on each other, or on covariates\index{covariate}.
The new book by \citet{HjortHMW}\index{Hjort, N. L.}\index{Holmes, C.}\index{Mueller, P.@M\"uller, P.} provides an excellent introduction and review of 
these developments.

\paragraph{P\'olya trees.} 
Ferguson's\index{Ferguson, T. S.} definition of the Dirichlet process focussed
on the (random) probabilities to be assigned to arbitrary
partitions\index{random partition} (Section \ref{sec:ferguson}). As we have
seen, the resulting distributions $G$ are almost surely discrete. An effective
way to modify this process to control continuity properties is to limit the
partitions to which elementary probabilities are assigned, and in the case of
P\'olya tree\index{Polya, G.@P\'olya, G.!P\'olya tree} processes this is
achieved by imposed a fixed binary partition of $\Omega$, and assigning
probabilities to successive branches in the tree through independent Beta
distributions\index{beta distribution|)}. The parameters of these distributions can be set to obtain various degrees of smoothness of the resulting $G$. This approach, essentially beginning with Ferguson himself, has been pursued by
\citet{Lavine92,Lavine94}\index{Lavine, M.}; see also
\citet{WalkerDLS}\index{Walker, S. G.|)}\index{Damien, P.|)}\index{Laud, P. W.|)}\index{Smith, A. F. M.|)}.

\section{P\'olya urn schemes and MCMC samplers} 
\label{sec:polya}

There is a huge literature on Markov chain Monte Carlo\index{Markov, A. A.!Markov chain Monte Carlo (MCMC)|(} methods for posterior sampling in
Dirichlet mixture models
\citep{MacEachern,EscobarW,MuellerEW,MacEachernM98,NealJCGS,
GreenR}\index{Escobar, M. D.}\index{West, M.}%
\index{Mueller, P.@M\"uller, P.}\index{Erkanli, A.}\index{Neal, R. M.}\index{Richardson, S.}.
Although these models have `variable dimension', the posteriors can be sampled
without necessarily using reversible jump methods\index{reversible jump method} \citep{Green}.

Cases where $G_0$ is not conjugate\index{conjugacy} to the data model $f(\cdot|\phi)$ demand keeping $\{\phi_i\}$ in the
state vector, to be handled through various augmentation or reversible jump schemes.
In the conjugate case, however, it is obviously appealing to integrate $\phi$
out, and target the Markov
chain on the posterior solely of the partition, generating $\phi$ values subsequently
as needed. To discuss this, we first go back to probability theory.

\subsection{The P\'olya urn representation of the Dirichlet process}

The P\'olya urn\index{Polya, G.@P\'olya, G.!P\'olya urn} is a simple and well-known discrete probability model for a reinforcement process: coloured balls are drawn sequentially from an urn; after each is drawn it is replaced, together with a new ball of the same \Index{colour}. This idea can be seen in a generalised form, in a recursive definition of the joint distribution of the $\{\phi_i\}$. 

Suppose that for each $n=0$, 1, 2, \ldots,
\bel{polyaurn}
\phi_{n+1}|\phi_1,\phi_2,\ldots,\phi_n \sim \frac{1}{n+\theta} \sum_{i=1}^n \delta_{\phi_i} +\frac{\theta}{n+\theta} G_0,
\ee
where $\theta>0$, $G_0$ is an arbitrary probability distribution, and
$\delta_\phi$ is a point probability mass at
$\phi$. \citet{BlackwellM}\index{Blackwell, D.}\index{MacQueen, J. B.} termed
such a sequence a P\'olya sequence\index{Polya, G.@P\'olya, G.!P\'olya sequence}; they showed that the conditional
distribution on the right hand side of (\ref{polyaurn}) converges to a random
probability distribution $G$ distributed as $DP(\theta,G_0)$, and that, given
$G$, $\phi_1$, $\phi_2$, \ldots\ are i.i.d. distributed as $G$. See also
\citet{Antoniak}\index{Antoniak, C. E.} and \citet{Pitman}\index{Pitman, J. [Pitman, J. W.]}.

Thus we have yet another approach to defining the Dirichlet process, at least in so far as specifying the joint distribution of the $\{\phi_i\}$ is concerned. This representation has a particular role, of central importance in computing inferences in DP models. This arises directly from (\ref{polyaurn}) and the \Index{exchangeability} of the $\{\phi_i\}$, for it follows that
\bel{polyaurn2}
\phi_i|\phi_{-i} \sim \frac{1}{n-1+\theta} \sum_{j\neq i} \delta_{\phi_j} +\frac{\theta}{n-1+\theta} G_0,
\ee
where $\phi_{-i}$ means $\{\phi_j:j=1,2,\ldots,n, j\neq i\}$. In this form,
the statement has an immediate role as the \emph{full conditional}
distribution for each component of $(\phi_i)_{i=1}^n$, and hence defines a
Gibbs sampler\index{Gibbs, J. W.!Gibbs sampler|(} update in a Markov chain Monte
Carlo\index{Markov, A. A.!Markov chain Monte Carlo (MCMC)|)} method aimed at
this target distribution. By conjugacy this remains true, with obvious changes
set out in the next section, for \index{sampling!posterior sampling|(}posterior sampling as well.

The P\'olya urn representation of the Dirichlet process has been the point of
departure for yet another class of probability models, namely
\index{sampling!species sampling}species sampling models
\citep{Pitman,Pitman96}, that are beginning to find a use
in statistical methodology \citep{IshwaranJ03}\index{Ishwaran, H.}\index{James, L. F.}. 

\subsection{The Gibbs sampler for posterior sampling of allocation variables}
\label{sec:postsamp}

We will consider posterior sampling\index{sampling!posterior sampling|)} in
the conjugate case in a more general setting, specialising back to the
Dirichlet process mixture case later. The set-up we will assume is based on a
partition model\index{partition!model}: it consists of a prior distribution $p(\cv|\theta)$ on
partitions $\cv$ of $\{1,2,\ldots,n\}$ with \index{hyperparameter|(}hyperparameter $\theta$, together
with a conjugate\index{conjugacy|(} model within each cluster. The prior on the cluster-specific
parameter $\phi_j$ has hyperparameter $\psi$, and is conjugate to the
likelihood\index{likelihood|(}, so that for any subset $C\subseteq \{1,2,\ldots,n\}$,
$p(Y_C|\psi)$ is known explicitly, where $Y_C$ is the subvector of
$(Y_i)_{i=1}^n$ with indices in $C$. We have
$$
p(Y_C|\psi) = \int \prod_{i\in C} p(Y_i|\phi) p(\phi|\psi) \,d\psi.
$$

We first consider only re-allocating a single item
at a time (the single-variable Gibbs
sampler\index{Gibbs, J. W.!Gibbs sampler|)} for $c_i$). Then repeatedly we withdraw an item, say $i$, from
the model, and reallocate it to a cluster according to the full
conditional\undex{full conditional distribution} for $c_i$, which is proportional to $p(\cv|Y,\theta,\psi)$. It is easy to see that we have two choices:
\BI
\item allocate $Y_i$ to a new cluster $C_\star$, with probability
$$
\propto p(\cv^{i\rightarrow \star}|\theta) \times p(Y_i|\psi),
$$
where $\cv^{i\rightarrow \star}$ denotes the current partition $\cv$
with $i$ moved to $C_\star$, or

\item allocate $Y_i$ to cluster $C_j^{-i}$, with probability
$$
\propto p(\cv^{i\rightarrow j}|\theta) \times
p(Y_{C_j^{-i}\cup\{i\}}|\psi)/ p(Y_{C_j^{-i}}|\psi),
$$
where $\cv^{i\rightarrow j}$ denotes the partition $\cv$, with $i$
moved to cluster $C_j$.
\EI

The ratio of marginal likelihoods $p(Y|\psi)$ in the second expression
can be interpreted as the
posterior predictive distribution\index{posterior predictive distribution} of $Y_i$ given those observations
already allocated to the cluster, i.e.
$p(Y_i|Y_{C_j^{-i}},\psi)$ (a multivariate $t$\index{multivariate $t$ distribution} for the Normal--inverse gamma\index{normal--inverse gamma model} set-up from Section \ref{sec:gene}).

For Dirichlet mixtures\index{Dirichlet, J. P. G. L.!Dirichlet mixture} we have, from (\ref{dppart}),
$$
p(\cv|\theta)=\frac{\Gamma(\theta)}{\Gamma(\theta+n)}
\theta^d\prod_{j=1}^d (n_j-1)!,
$$
where $n_j=\#C_j$ and $\cv=(C_1,C_2,\ldots,C_d)$, so the 
re-allocation probabilities are explicit and simple in form.

But the same sampler can be used for many other partition models, and the idea is not limited to moving one item at a time.

\subsection{When the P\'olya urn sampler applies}
All we require of the model for the P\'olya urn
sampler\index{Polya, G.@P\'olya, G.!P\'olya urn sampler} to be available for
posterior simulation\index{simulation!posterior simulation} are that 
\begin{enumerate}[1.]
\item a partition $\cv$ of $\{1,2,\ldots,n\}$ is drawn from a prior distribution with parameter $\theta$;
\item conditionally on $\cv$, parameters $(\phi_1,\phi_2,\ldots,\phi_d)$ are drawn independently from a distribution $G_0$ (possibly with a hyperparameter\index{hyperparameter|)} $\psi$);
\item conditional on $\cv$ and on $\phi=(\phi_1,\phi_2,\ldots,\phi_d)$,
$\{y_1,y_2,\ldots,y_n\}$ are drawn independently, from not necessarily identical distributions\break $p(y_i|\cv,\phi)=f_i(y_i|\phi_j)$ for $i\in C_j$, for which $G_0$ is conjugate\index{conjugacy|)}.
\end{enumerate}
If these all hold, then the P\'olya urn sampler can be used; we see from Section \ref{sec:postsamp} that it will involve computing only marginal likelihoods, and ratios of the partition prior, up to a multiplicative constant. The first factor depends only on $G_0$ and the likelihood\index{likelihood|)}, the second only on the partition model\index{partition!model|)}.

\paragraph{Examples.}

$p(\cv^{i\rightarrow \star}|\theta)$ and
$p(\cv^{i\rightarrow j}|\theta)$
are proportional simply to
\BI
\I $\theta$ and $\#C_j^{-i}$ for the DP mixture model\index{Dirichlet, J. P. G. L.!Dirichlet process mixture model},
\I $(k-d(\cv^{-i}))\delta$ and $\#C_j^{-i}+\delta$ for the Dirichlet--multinomial 
finite mixture
model\index{Dirichlet, J. P. G. L.!Dirichlet multinomial finite mixture model@Dirichlet--multinomial finite mixture model},
\I $\theta+\alpha d(\cv^{-i})$ and $\#C_j^{-i}-\alpha$ for the
Kingman--Pitman--Yor\index{Kingman, J. F. C.}\index{Pitman, J. [Pitman, J. W.]}\index{Yor, M.} two-paramet\-er Poisson--Dirichlet process\index{two-parameter Poisson--Dirichlet process} (Section \ref{sec:general}).
\EI

It is curious that the ease of using the P\'olya urn sampler has often been cited as motivation to use Dirichlet process mixture models, when the class of models for which it is equally readily used is so wide.

\subsection{Simultaneous re-allocation}\index{simultaneous re-allocation}
There is no need to restrict to updating only one $c_i$
at a time: the idea extends to simultaneously re-allocating
any subset of items \emph{currently in the same cluster}.

The notation can be rather cumbersome, but again the subset
forms a new cluster, or moves to an existing cluster,
with relative probabilities that are each products
of two terms:
\BI
\I the relative (new) partition prior probabilities, and
\I the \Index{predictive density} of the moved set of
item data, given those already in the receiving cluster.
\EI

A more sophisticated variant on this scheme has been proposed by
\citet{NobileF}\index{Nobile, A.}\index{Fearnside, A. T.}, and studied in the case
of finite mixture models\index{mixture model}.

\section{A coloured Dirichlet process}
\label{sec:cdp}\index{cluster(ing)|(}

For the remainder of this note, we focus on the use of these models for
clustering, rather than \Index{density estimation} or other kinds of inference. There
needs to be a small caveat---mixture models are commonly used either for
clustering, or for fitting non-standard distributions; in a problem demanding
\emph{both}, we cannot expect to be able meaningfully to identify clusters
with the components of the mixture, since multiple components may be needed to
fit the non-standard distributional shape within each cluster. Clustered
Dirichlet process methodology in which there is clustering at two levels that
can be used for such a purpose is under development by Dan Merl\index{Merl,
  D.} and Mike West\index{West, M.}
at Duke\index{Duke University} (personal communication).

Here we will not pursue this complication, and simply consider a \Index{mixture model} used for clustering in the obvious way. 

In many domains of application, practical considerations suggest that the
clusters in the data do not have equal standing; the most common such
situation is where there is believed to be a `background' cluster, and one or
several `foreground' clusters, but more generally, we can imagine there being
several classes of cluster, and our prior beliefs are represented by the idea
that cluster labels are exchangeable within these classes, but not overall. It
would be common, also, to have different beliefs about cluster-specific
parameters within each of these classes.

In this section, we present a variant on standard mixture/cluster models of
the kinds we have already discussed, aimed at modelling this situation of
partial exchangeability\index{exchangeability!partial exchangeability} of
cluster labels. We stress that it will remain true
that, \emph{a priori}, item labels are exchangeable, and that we have no prior
information that particular items are drawn to particular classes of cluster;
the analysis is to be based purely on the data $\{Y_i\}$.

We will describe the class of a cluster henceforth as its `colour'\index{colour|(}. To define a variant on the DP in which not all clusters are exchangeable:
\BE[1.]
\I for each `colour' $k=1$, 2, \ldots, 
draw $G_k$ from
a Dirichlet process DP$(\theta_k,G_{0k})$, independently for each $k$;
\I draw weights $(w_k)$ from the Dirichlet
distribution\index{Dirichlet, J. P. G. L.!Dirichlet distribution|(}
Dir$(\gamma_1,\gamma_2,\ldots)$, independently of the $G_k$;
\I define $G$ on $\{k\}\times \Omega$ by $G(k,B)=w_k G_k(B)$;
\I draw colour--parameter pairs $(k_i,\phi_i)$ i.i.d.\ from $G$, $i=1$, 2,
\ldots, $n$.
\EE

This process, denoted CDP$(\{(\gamma_k,\theta_k,G_{0k})\})$, is a Dirichlet
mixture\index{Dirichlet, J. P. G. L.!Dirichlet mixture} of Dirichlet processes
(with different base measures)\index{base measure}, $\sum_k w_k \mbox{DP}(\theta_k,G_{0k})$, with
the added feature that the the colour of each cluster is identified (and
indirectly observed), while labelling of clusters within colours is arbitrary.

It can be defined by a `stick-breaking-and-colouring'\index{stick breaking}\index{colouring} construction:
\BE[1.]
\I colour segments of the stick using the Dirichlet$(\{\gamma_k\})$-distributed weights;
\I break each coloured segment using an infinite sequence of independent
Beta$(1,\theta_k)$\index{beta distribution} variables $V_{jk}$;
\I draw $\phi^\star_{jk}\sim G_{0k}$, i.i.d., $j=1$, 2, \ldots; $k=1$, 2, \ldots;
\I define $G_k$ to be the discrete distribution putting
probability $(1-V_{1k})(1-V_{2k})\cdots(1-V_{j-1,k})V_{jk}$
on $\phi^\star_{jk}$.
\EE

Note that in contrast to other elaborations to more structured data of the Dirichlet process model, in which the focus is on nonparametric analysis and more sharing of information would be desirable, here, where the focus is on clustering, we are content to leave the atoms and weights within each colour completely uncoupled \emph{a priori}.

\subsection{Coloured partition distribution} 
The \Index{coloured Dirichlet process (CDP)} generates the following partition model\index{partition!model}: partition $\{1,2,\ldots,n\}=\bigcup_k\bigcup_{j=1}^{d_k} C_{kj}$
at random, where $C_{kj}$ is the $j$th cluster of colour $k$, so that
$$
p(C_{11},C_{12},\ldots,C_{1d_1};C_{21},\ldots,C_{2d_2};C_{31},\ldots) = 
$$
$$
\frac{\Gamma(\sum_k\gamma_k)}{\Gamma(n+\sum_k\gamma_k)}
\prod_k \left(
\frac{\Gamma(\theta_k) \Gamma(n_k+\gamma_k)}
{\Gamma(n_k+\theta_k)\Gamma(\gamma_k)}
\theta_k^{d_k}\prod_{j=1}^{d_k} (n_{kj}-1)! \right),
$$
where $n_{kj}=\#C_{kj}$, $n_k=\sum_j n_{kj}$.

It is curious to note that this expression simplifies when $\theta_k \equiv
\gamma_k$, although such a choice seems to have no particular significance in
the probabilistic construction of the model. Only when it is also true that
the $\theta_k$ are independent of $k$ (and the colours are ignored) does the
model degenerate to an ordinary Dirichlet
process.

The clustering remains exchangeable\index{exchangeability} over items. 
To complete the construction of the model, analogously to Section \ref{sec:partition}, for $i \in C_{kj}$, we set $k_i=k$ and $\phi_i=\phi^\star_j$, where $\phi^\star_j$ are drawn i.i.d. from $G_{0k}$.

\subsection{P\'olya urn sampler for the CDP}
The explicit availability of the (coloured) partition distribution\index{partition!distribution}
immediately allows generalisation of the
P\'olya-urn\index{Polya, G.@P\'olya, G.!P\'olya urn sampler|(}
\index{Gibbs, J. W.!Gibbs sampler}Gibbs sampler 
to the CDP.

In reallocating item $i$, let $n_{kj}^{-i}$ denote the number \emph{among the remaining items} currently allocated to $C_{kj}$,
and define $n_k^{-i}$ accordingly. Then reallocate $i$ to
\BI
\I a new cluster of colour $k$, with probability $\propto 
\theta_k\times (\gamma_k+n_k^{-i})/(\theta_k+n_k^{-i})\times p(Y_i|\psi)$, for
$k=1$, 2, \ldots;
\I the existing cluster $C_{kj}$, with probability $\propto 
n_{kj}^{-i}\times (\gamma_k+n_k^{-i})/(\theta_k+n_k^{-i})\times
p(Y_i|Y_{C_{kj}^{-i}},\psi)$, for $j=1$, 2, \ldots, $n_k^{-i}$; $k=1$, 2, \ldots.
\EI
 
Again, the expressions simplify when $\theta_k \equiv \gamma_k$.

\subsection{A Dirichlet process mixture with a background cluster}


In many applications of probabilistic clustering, including the gene
expression\index{gene!expression} example from Section \ref{sec:gene}, it is natural to suppose a
`background' cluster\undex{cluster(ing)!background cluster} that is not 
\emph{a priori} exchangeable\index{exchangeability} with the others. 
One way to think about this is to adapt the `limit of finite mixtures' view from Section \ref{sec:finitemixture}: 
\BE[1.]
\I draw $(w_0,w_1,w_2,\ldots,w_k)\sim\mbox{Dirichlet}(\gamma,\delta,\ldots,\delta)$;\index{Dirichlet, J. P. G. L.!Dirichlet distribution|)}
\I draw $c_i\in\{0,1,\ldots,k\}$ with $P\{c_i=j\}=w_j$, i.i.d., $i=1$, \ldots,
$n$;
\I draw $\phi^\star_0\sim H_0$, $\phi^\star_j\sim G_0$, i.i.d., $j=1$, \ldots,
$k$;
\I set $\phi_i=\phi^\star_{c_i}$.
\EE
Now let $k\to\infty$, $\delta\to 0$ such that $k\delta\to \theta$,
but leave $\gamma$ fixed. The cluster labelled 0 represents the `background'.

The background cluster model\index{cluster(ing)!background cluster model|(} is a special case of the CDP, specifically
CDP$(\{(\gamma,0,H_0),(\theta,\theta,G_0)\})$. The two colours correspond to the background and regular clusters. 
The limiting-case DP$(0,H_0)$
is a point mass, randomly drawn from $H_0$. 
We can go a little further in a regression\index{regression|(} setting, and allow different regression models for each colour.\index{colour|)}

The P\'olya urn sampler\index{Polya, G.@P\'olya, G.!P\'olya urn sampler|)} for prior or posterior simulation is readily adapted.
When re-allocating item $i$, there are three kinds of choice:
a new cluster $C_\star$, the `top table' $C_0$, or a regular cluster $C_j,j\neq 0$:
the corresponding prior probabilities
$p(\cv^{i\rightarrow \star}|\theta)$,
$p(\cv^{i\rightarrow 0}|\theta)$ and
$p(\cv^{i\rightarrow j}|\theta)$
are proportional to
$\theta$, $(\gamma+\#C_0^{-i})$ and $\#C_j^{-i}$ 
for the background cluster CDP model.

\subsection{Using the CDP in a clustered regression model}
\label{sec:clusreg}

As a practical illustration of the use of the CDP background cluster
model,\index{Dirichlet, J. P. G. L.!Dirichlet process (DP)|)}
we discuss a regression set-up that expresses a vector of 
measurements $\mathbf{y}_{i}=(y_{i1},\ldots,y_{iS})$ for $i=1$, \ldots, $n$,
where $S$ is the number of samples,
as a linear combination of known covariates\index{covariate|(},
$(\mathbf{z}_1\cdots\mathbf{z}_S)$ with dimension
$K^{\prime}$ and $(\mathbf{x}_1\cdots\mathbf{x}_S)$
with dimension $K$.
These two collections of covariates, and the corresponding 
regression coefficients $\mbox{\boldmath{$\delta$}}_j$ and 
$\mbox{\boldmath{$\beta$}}_j$, are distinguished since we wish to hold one
set of regression coefficients fixed in the background cluster.
We assume
\begin{eqnarray} \label{Regression Model 2} \mathbf{y}_{i}
=\left[\begin{array}[c]{c}y_{i1}\\\vdots\\y_{iS}\end{array}\right]%
&=&\sum_{k^{\prime}=1}^{K^{\prime}}\delta_{jk^{\prime}}\left[\begin{array}[c]{c}z_{1k^{\prime}}\\\vdots\\z_{Sk^{\prime}}\end{array}\right]+%
\sum_{k=1}^{K}\beta_{jk}\left[\begin{array}[c]{c}x_{1k}\\\vdots\\x_{Sk}\end{array}\right]%
+\left[\begin{array}[c]{c}\epsilon_{j1}\\\vdots\\\epsilon_{jS}\end{array}
\right]\nonumber\\%
&=&
[\mathbf{z}_1\cdots\mathbf{z}_S]^{\prime}{\mbox{\boldmath{$\delta$}}_j}
+[\mathbf{x}_1\cdots\mathbf{x}_S]^{\prime}{\mbox{\boldmath{$\beta$}}_j}
+\mbox{\boldmath{$\epsilon$}}_{j}%
\end{eqnarray}
where $\mbox{\boldmath{$\epsilon$}}_{j}\sim
N(\mathbf{0}_{S\times{1}},\tau_{j}^{-1}\mathbf{I}_{S\times{S}})$,
$\mathbf{0}_{S\times{1}}$ is the $S$--dimension zero vector and
$\mathbf{I}_{S\times{S}}$ is the order--$S$ identity matrix. 
Here, $\mbox{\boldmath{$\delta$}}_j$, $\mbox{\boldmath{$\beta$}}_j$
and $\tau_{j}$ are cluster-specific parameters. The profile of measurements for individual $i$ is
$\mathbf{y}_i=[y_{i1}\cdots y_{iS}]^{\prime}$ for $i=1$, \ldots, $n$. 
Given the covariates $\mathbf{z}_s=[z_{s1}\cdots
z_{sK^{\prime}}]^{\prime}$, $\mathbf{x}_s=[x_{s1}\cdots
x_{sK}]^{\prime}$, and the cluster $j$, the parameters/\Index{latent
variables} are
${\mbox{\boldmath{$\delta$}}_j}=[\delta_{j1}\cdots\delta_{jK^{\prime}}]^{\prime}$
,
${\mbox{\boldmath{$\beta$}}_j}=[\beta_{j1}\cdots\beta_{jK}]^{\prime}$
and $\tau_j$. The kernel is now represented as
$k(\mathbf{y}_i\vert\mbox{\boldmath{$\delta$}}_j,\mbox{\boldmath{$\beta$}}_j,\tau_j)$,
which is a multivariate Normal density, $N(
[\mathbf{z}_1\cdots\mathbf{z}_S]^{\prime}{\mbox{\boldmath{$\delta$}}_j}
+[\mathbf{x}_1\cdots\mathbf{x}_S]^{\prime}{\mbox{\boldmath{$\beta$}}_j}
,\tau_{j}^{-1}\mathbf{I}_{S\times{S}})$. In particular, we take
different probability measures, the parameters of heterogeneous DP, for the background and regular clusters,
\begin{align*}
\mathbf{u}_0 = (\mbox{\boldmath{$\delta$}}_0,\mbox{\boldmath{$\beta$}}_0,\tau_0)
&\sim H_0(d\mbox{\boldmath{$\delta$}}_0,d\mbox{\boldmath{$\beta$}}_0,d\tau_0)\\
&= \delta_{\mbox{\boldmath{$\delta$}}_0} (d\mbox{\boldmath{$\delta$}}_0)\times \mbox{Normal--Gamma} (d\mbox{\boldmath{$\beta$}}_0,d\tau_0^{-1});\\
\mathbf{u}_j = (\mbox{\boldmath{$\delta$}}_j,\mbox{\boldmath{$\beta$}}_j,\tau_j)
&\sim G_0(d\mbox{\boldmath{$\delta$}}_j,d\mbox{\boldmath{$\beta$}}_j,d\tau_j)\\*
&= \mbox{Normal--Gamma} (d(\mbox{\boldmath{$\delta$}}_j^{\prime},\mbox{\boldmath{$\beta$}}_j^{\prime})^\prime,d\tau_j^{-1})\\*
 &\qquad\qquad\text{ for $j = 1$, \ldots, $n(\mathbf{p})-1$}.
\end{align*}
Here $H_0$ is a probability measure that includes a point mass at
$\mbox{\boldmath{$\delta$}}_0$ and a Normal--Gamma\index{gamma distribution} density for
$\mbox{\boldmath{$\beta$}}_0$ and $\tau_0^{-1}$. On the other hand,
we take $G_0$ to be a probability measure that is a Normal--Gamma
density for
$(\mbox{\boldmath{$\delta$}}_j^{\prime},\mbox{\boldmath{$\beta$}}_j^{\prime})^\prime$
and $\tau_j^{-1}$. Thus the regression\index{regression|)} parameters corresponding to the $z$ covariates\index{covariate|)}
are held fixed at $\mbox{\boldmath{$\delta$}}_0$ in the background cluster, but not in the others.

We will first discuss the marginal distribution for the regular
clusters. Given $\tau_{j}$,
$(\mbox{\boldmath{$\delta$}}_j^{\prime},\mbox{\boldmath{$\beta$}}_j^{\prime})^{\prime}$
 follows the ($K^{\prime}+K$)--dimensional multivariate Normal with mean
$\widetilde{\mathbf{m}}$ and variance
$(\tau_j\widetilde{\mathbf{t}})^{-1} $ and $\tau_j$ follows the
univariate Gamma with shape $\widetilde{a}$ and scale
$\widetilde{b}$. We denote the joint distribution $G_0(d
(\mbox{\boldmath{$\delta$}}_j^{\prime},\mbox{\boldmath{$\beta$}}_j^{\prime})^{\prime}
,d\tau_j)$ as a joint Gamma and Normal distribution,
$\mbox{Normal--}\allowbreak\mbox{Gamma}(\widetilde{a},\widetilde{b},\widetilde{\mathbf{m}},\widetilde{\mathbf{t}})
$, and further we take
\begin{equation}
\widetilde{\mathbf{m}} =
\begin{bmatrix}\widetilde{\mathbf{m}}_\delta \\ \widetilde{\mathbf{m}}_\beta\end{bmatrix}
\text{ and } \widetilde{\mathbf{t}} =
\begin{bmatrix}\widetilde{\mathbf{t}}_\delta & 0 \\ 0 & \widetilde{\mathbf{t}}_\beta\end{bmatrix}.
\end{equation}
Based on this set-up, we have 
\begin{multline}
m_{G_0}(\mathbf{y}_{C_{j}})=%
\\ t_{2\widetilde{a}}(\mathbf{Y}_{C_j}\vert%
\mathbf{Z}_{C_j}\widetilde{\mathbf{m}}_\delta + \mathbf{X}_{C_j} \widetilde{\mathbf{m}}_\beta%
,%
\frac{\widetilde{b}}{\widetilde{a}} (
\mathbf{Z}_{C_j}\widetilde{\mathbf{t}}_\delta^{-1}\mathbf{Z}_{C_j}^{\prime} +%
\mathbf{X}_{C_j}\widetilde{\mathbf{t}}_\beta^{-1} \mathbf{X}_{C_j}^{\prime}%
+\mathbf{I}_{e_jS\times e_jS})),\label{marginal for regular table}%
\end{multline}
where
$\mathbf{Y}_{C_{j}}=[\mathbf{y}_{i_1}^{\prime}\cdots\mathbf{y}_{i_{e_j}}^{\prime}]^{\prime}$,
$\mathbf{X}_{C_{j}}=[[\mathbf{x}_1\cdots\mathbf{x}_S]\cdots[\mathbf{x}_1\cdots\mathbf{x}_S]]^{\prime}$
and
$\mathbf{Z}_{C_{j}}=[[\mathbf{z}_1\cdots\mathbf{z}_S]\cdots[\mathbf{z}_1\cdots\mathbf{z}_S]]^{\prime}$
for $C_j=\{i_{1},\ldots,i_{e_j}\}$. Note that $\mathbf{Y}_{C_{j}}$
is a $e_jS$ vector, $\mathbf{Z}_{C_{j}}$ is a $e_jS\times
K^{\prime}$  matrix and $\mathbf{X}_{C_{j}}$ is a $e_jS\times K$
matrix. Moreover, $m_{G_0}(\mathbf{y}_{C_{j}})$ is a multivariate
$t$\index{multivariate $t$ distribution|(} density with mean
$\mathbf{Z}_{C_j}\widetilde{\mathbf{m}}_\delta + \mathbf{X}_{C_j} \widetilde{\mathbf{m}}_\beta$%
, scale
$$\frac{\widetilde{b}}{\widetilde{a}} (
\mathbf{Z}_{C_j}\widetilde{\mathbf{t}}_\delta^{-1}\mathbf{Z}_{C_j}^{\prime} +%
\mathbf{X}_{C_j}\widetilde{\mathbf{t}}_\beta^{-1} \mathbf{X}_{C_j}^{\prime}%
+\mathbf{I}_{e_jS\times e_jS}))$$
 and degree of freedom
$2\widetilde{a}$.

For the background cluster\index{cluster(ing)!background cluster model|)}, we take $H_0$ to be a joint Gamma and
Normal distribution,
$\mbox{Normal--Gamma}(\overline{a},\overline{b},\overline{\mathbf{m}}_\beta,\overline{\mathbf{t}}_\beta)$.
The precision $\tau_0$ follows the univariate Gamma with shape
$\overline{a}$ and scale $\overline{b}$. Given $\tau_{0}$,
$\mbox{\boldmath{$\beta$}}_0$
 follows the $K$--dimension multivariate Normal with mean
$\overline{\mathbf{m}}_\beta$ and variance
$(\tau_0\overline{\mathbf{t}}_\beta)^{-1}$ and $\tau_0$ follows the
univariate Gamma with shape $\overline{a}$ and scale $\overline{b}$.
The marginal distribution becomes
\begin{equation} m_{H_0}(\mathbf{y}_{C_{0}})=%
t_{2\overline{a}}(\mathbf{Y}_{C_j}\vert \mathbf{Z}_{C_j} \mbox{\boldmath{$\delta$}}_0 +\mathbf{X}_{C_j} \overline{\mathbf{m}}_\beta%
, \frac{\overline{b}}{\overline{a}}
(\mathbf{X}_{C_j}\overline{\mathbf{t}}_\beta^{-1}
\mathbf{X}_{C_j}^{\prime} +\mathbf{I}_{e_jS\times e_jS}) )\label{marginal for top table}.
\end{equation}
So $m_{H_0}(\mathbf{y}_{C_{0}})$ is a multivariate $t$
density\index{multivariate $t$ distribution|)} with
mean $\mathbf{Z}_{C_j} \mbox{\boldmath{$\delta$}}_0 +\mathbf{X}_{C_j} \overline{\mathbf{m}}_\beta$%
, scale
$$\frac{\overline{b}}{\overline{a}}
(\mathbf{X}_{C_j}\overline{\mathbf{t}}_\beta^{-1}
\mathbf{X}_{C_j}^{\prime} +\mathbf{I}_{e_jS\times e_jS})$$
and degree
of freedom $2\overline{a}$.

In some applications, the $x$s and $\beta$s are not needed and so can be omitted, 
and we consider the following model,
\begin{equation}\label{Regression Model 1} \mathbf{y}_{i}
=\left[\begin{array}[c]{c}y_{i1}\\\vdots\\y_{iS}\end{array}\right]%
=\sum_{k^{\prime}=1}^{K^{\prime}}\delta_{jk^{\prime}}\left[\begin{array}[c]{c}z_{1k^{\prime}}\\\vdots\\z_{Sk^{\prime}}\end{array}\right]%
+\left[\begin{array}[c]{c}\epsilon_{j1}\\\vdots\\\epsilon_{jS}\end{array}
\right]%
=
[\mathbf{z}_1\cdots\mathbf{z}_S]^{\prime}{\mbox{\boldmath{$\delta$}}_j}
+\mbox{\boldmath{$\epsilon$}}_{j};
\end{equation}
here we assume that $K=0$ or
$[\mathbf{x}_1\cdots\mathbf{x}_S]^{\prime}=\mathbf{0}_{S\times K}$
where $\mathbf{0}_{S\times K}$ is the $S \times K$ matrix with all
zero entries of the model (\ref{Regression Model 2}). We can derive
the marginal distributions analogous to (\ref{marginal for regular
table}) and (\ref{marginal for top table}),
\begin{eqnarray}
m_{G_0}(\mathbf{y}_{C_{j}})&=&%
t_{2\widetilde{a}}(\mathbf{Y}_{C_j}\vert%
\mathbf{Z}_{C_j}\widetilde{\mathbf{m}}_\delta,%
\frac{\widetilde{b}}{\widetilde{a}} (
\mathbf{Z}_{C_j}\widetilde{\mathbf{t}}_\delta^{-1}\mathbf{Z}_{C_j}^{\prime}%
+\mathbf{I}_{e_jS\times e_jS})),\\%
 m_{H_0}(\mathbf{y}_{C_{0}})&=&%
t_{2\overline{a}}(\mathbf{Y}_{C_j}\vert \mathbf{Z}_{C_j}
\mbox{\boldmath{$\delta$}}_0 , \frac{\overline{b}}{\overline{a}}
\mathbf{I}_{e_jS\times e_jS}).
\end{eqnarray}
Here
$t_{\nu}\left(\mathbf{x}\left\vert\mbox{\boldmath{$\mu$}},\mbox{\boldmath{$\Sigma$}}\right.\right)$
is a multivariate $t$ density in $d$ dimensions with mean
$\mbox{\boldmath{$\mu$}}$ and scale $\mbox{\boldmath{$\Sigma$}}$
with degrees of freedom $\nu$;
\begin{equation}
t_{\nu}(\mathbf{x}\vert\mbox{\boldmath{$\mu$}},\mbox{\boldmath{$\Sigma$}})
= \dfrac{\Gamma((\nu+d)/2)}{\Gamma((\nu)/2)}\dfrac{\vert\mbox{\boldmath{$\Sigma$}}\vert ^{-1/2}}{(\nu\pi)^{d/2}}%
(1+\dfrac{1}{\nu}(\mathbf{x}-\mbox{\boldmath{$\mu$}})^{\prime}\mbox{\boldmath{$\Sigma$}}^{-1}(\mathbf{x}-\mbox{\boldmath{$\mu$}}))^{-(\nu+d)/2}.
\end{equation}

\begin{figure}[htbp]
\begin{center}
\resizebox{90mm}{!}{\includegraphics{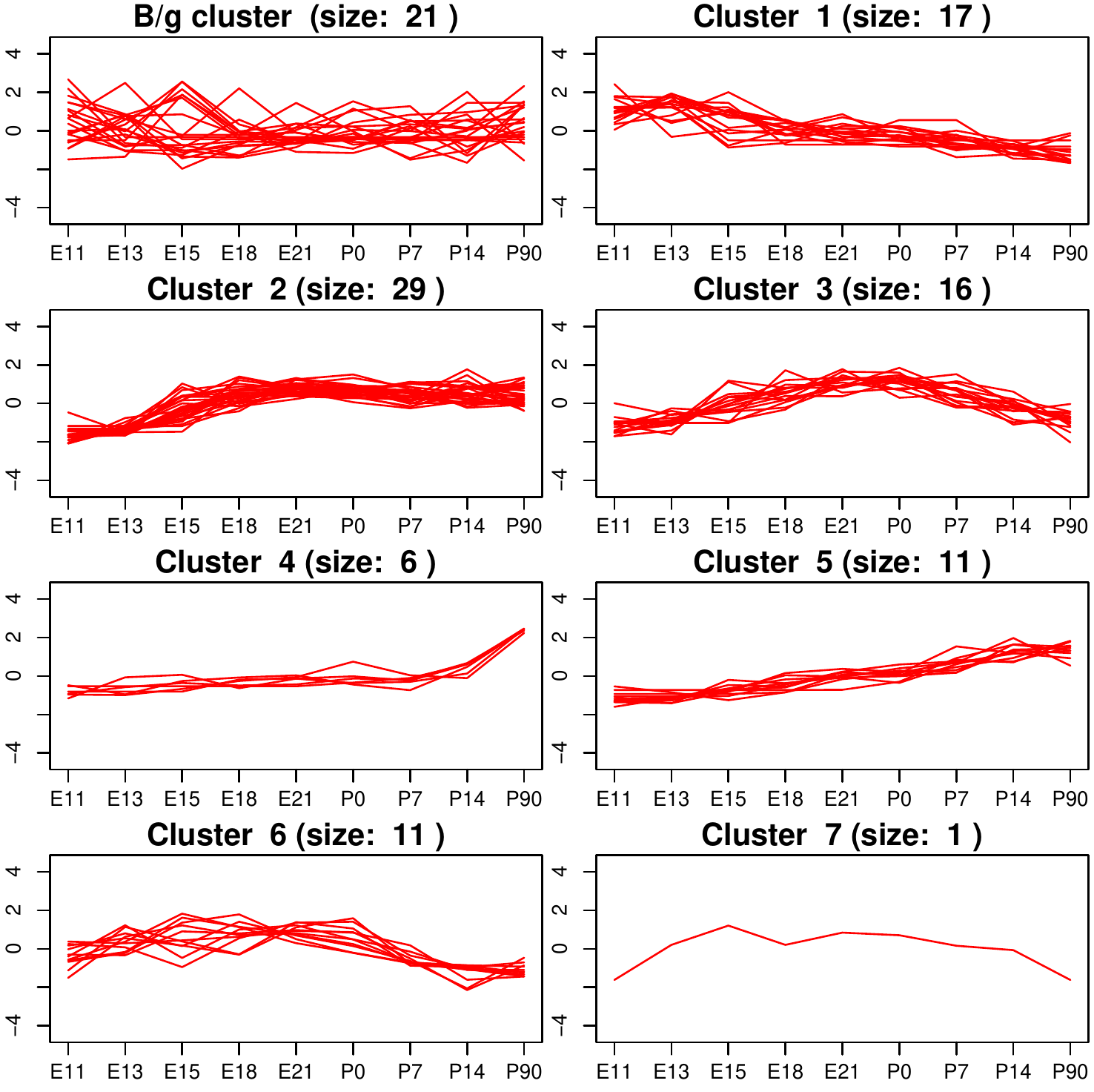}}
\caption{Profile plot\index{profile plot} of our partition estimate for the Rat data set of
\citet{Wen}.}
\label{fig:profiles}
\end{center}
\end{figure}

\subsection{Time-course gene expression data}
\label{sec:timecourse}\index{data!rat time-course experiment|(}

We consider the application of this methodology to data from a gene
expression\index{gene!expression}
\undex{data!time-course experiment}time-course experiment.
\citet{Wen}\index{Wen, X.|(} studied the \Undex{central nervous system} development of the
\Index{rat}; see also \citet{Yeung}\index{Teung, K. Y.}\index{Haynor,
  D. R.}\index{Ruzzo, W. L.}. The \Index{mRNA} expression levels of 112 genes
were recorded over the period of development of the central nervous system. In
the dataset, there are 9 records for each gene over 9 time
points; they are from embryonic days 11, 13, 15, 18, 21, postnatal
days 0, 7, 14, and the `adult' stage (postnatal day 90). 

In their analysis, \citet{Wen}
obtained 5 clusters/waves (totally 6 clusters), taken to
characterize distinct phases of development. The data set is
available at
\url{http://faculty.washington.edu/kayee/cluster/GEMraw.txt}. We
take $S=9$ and $K^{\prime}=5$. The design matrix of covariates\index{covariate} is taken
to be \begin{equation}
[\mathbf{z}_1\cdots\mathbf{z}_S]^{\prime}=\left[
  \begin{array}{ccccccccc}
    1 & 1 & 1 & 1 & 1 & 0 & 0 & 0 & 0 \\
    11 & 13 & 15 & 18 & 21 & 0 & 0 & 0 & 0 \\
    0 & 0 & 0 & 0 & 0 & 1 & 1 & 1 & 0 \\
    0 & 0 & 0 & 0 & 0 & 0 & 7 & 14 & 0 \\
    0 & 0 & 0 & 0 & 0 & 0 & 0 & 0 & 1
  \end{array}
\right]^{\prime},\nonumber%
\end{equation} 
representing piecewise linear dependence on time, within three separate
phases (embryonic, postnatal and adult).

In our analysis of these data, we
take $\theta=1$, $\gamma=5$, $\widetilde{a}=\overline{a}=0.01$,
$\widetilde{b}=\overline{b}=0.01$,
$\widetilde{\mathbf{m}}_\delta=\overline{\mathbf{m}}_\delta=[0\cdots0]^{\prime}$,
$\widetilde{\mathbf{t}}_\delta=\overline{\mathbf{t}}_\delta=0.01\mathbf{I}$,
$\widetilde{\mathbf{m}}_\beta=[0\cdots0]^{\prime}$,
$\widetilde{\mathbf{t}}_\beta=0.01\mathbf{I}$ and
$\mbox{\boldmath{$\delta$}}_0=[0\cdots0]^{\prime}$. %
The P\'olya urn sampler\index{Polya, G.@P\'olya, G.!P\'olya urn sampler} was implemented, and run for 20000 sweeps starting from the partition consisting of all singleton
clusters, 10000 being discarded as \Index{burn-in}. We then use the
last 10000 partitions sampled as in \citet{LauG}\index{LauJW@Lau, J. W.}, to
estimate the optimal Bayesian partition\index{Bayes, T.!Bayesian partition} on a decision-theoretic basis, using a pairwise coincidence \Index{loss function} that equally weights false `positives' and `negatives'.

We present some views of the resulting posterior analysis of this data set.

\begin{figure}[htbp]
\begin{center}
\resizebox{90mm}{!}{\includegraphics{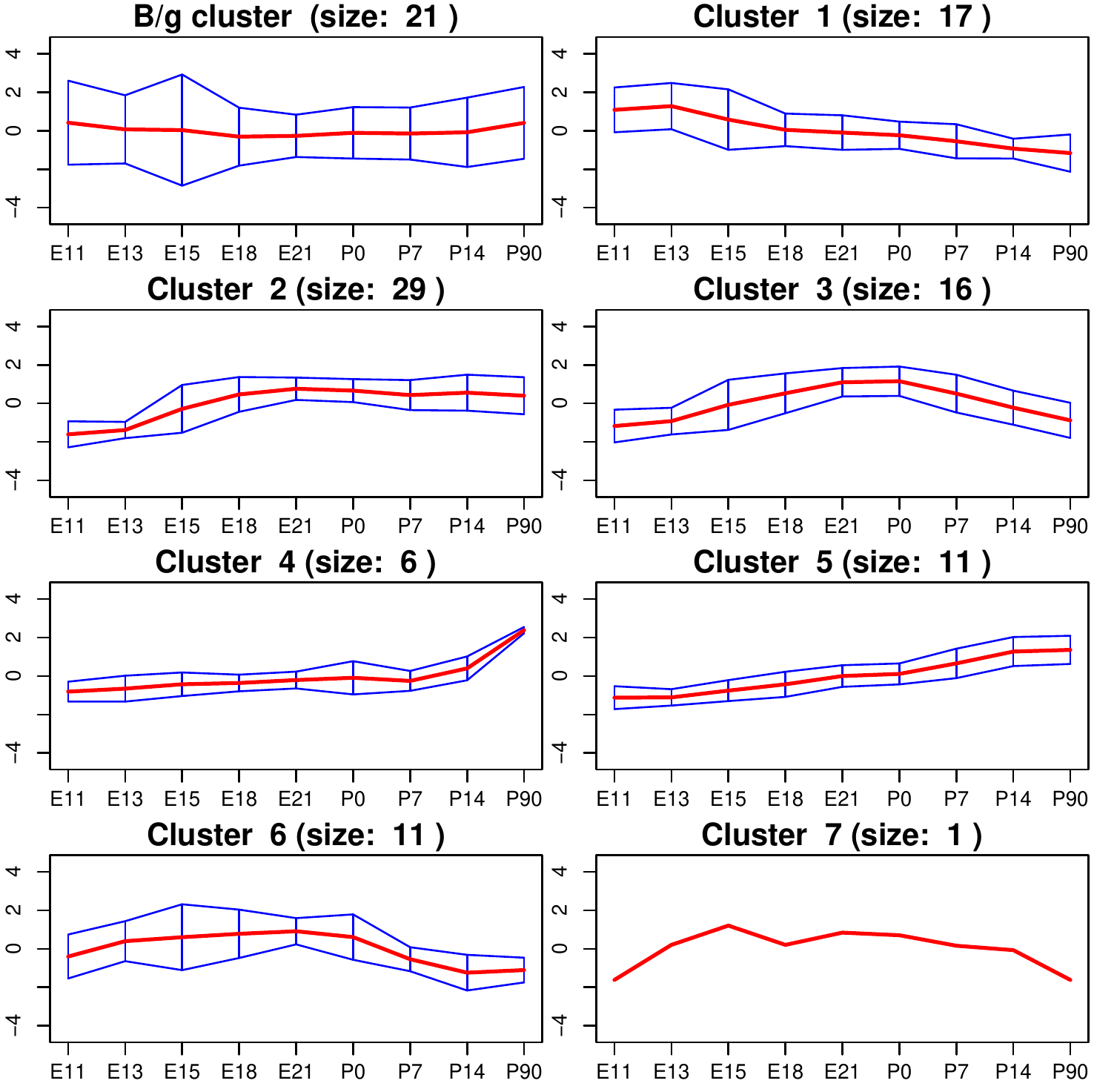}}
\caption{Mean and 95\% CI of genes
across clusters of our partition estimate.}
\label{fig:summary}
\end{center}
\end{figure}

\begin{figure}[htbp]
\begin{center}
\resizebox{109mm}{!}{\includegraphics{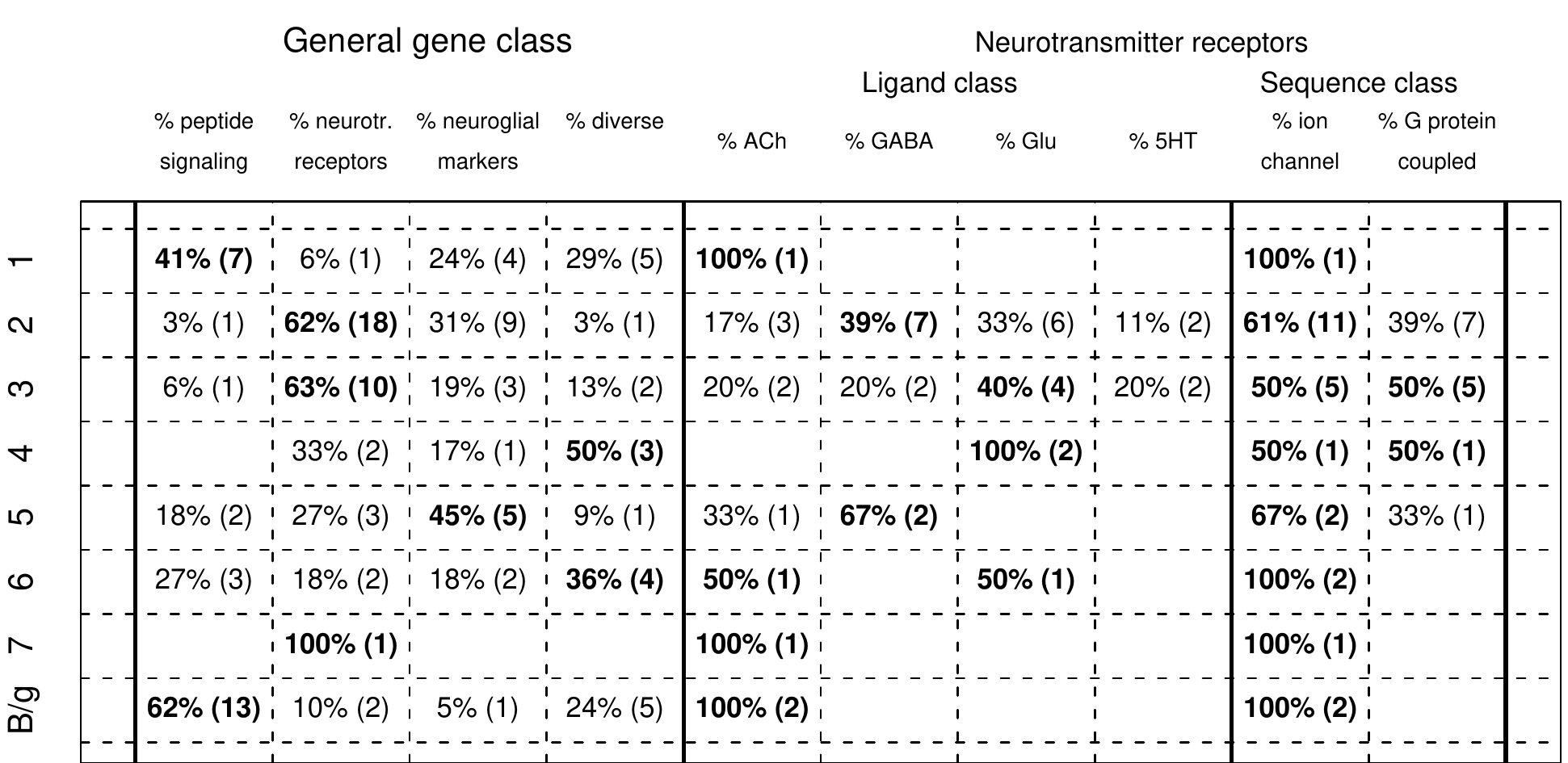}}
\caption{Biological functions of our Bayesian partition estimate for the genes in the data set of
\citet{Wen}, showing correspondence between inferred clusters and the functional categories of the genes.
All genes are classified into 4 general
\index{gene!class}gene classes. Additionally, the
Neurotransmitter genes have been further categorised by 
ligand class and functional sequence class.
Boldface type represents the dominant class in the cluster, in each categorisation.}
\label{fig:function}
\end{center}
\end{figure}

Figure \ref{fig:profiles} shows the profiles in the inferred clusters plotted, 
and Figure \ref{fig:summary} the
mean and the 95\% CI of the clusters. Figure \ref{fig:function}
cross-tabulates the clusters with the biological functions attributed to the relevant
genes by \citet{Wen}\index{Wen, X.|)}.\index{cluster(ing)|)}\index{data!rat time-course experiment|)}

\paragraph{Acknowledgements}

I am grateful to Sonia Petrone\index{Petrone, S.} and Simon
Tavar\'e\index{Tavar\'e, S.} for some pointers to the literature, John
Lau\index{LauJW@Lau, J. W.} for the analysis of the gene expression data, and John
Kingman\index{Kingman, J. F. C.!influence} for his undergraduate lectures in Measure and Probability.

\bibliography{dirichlet}

\providecommand\SortNoop[1]{}
\begin{thebibliography}{52}
\expandafter\ifx\csname natexlab\endcsname\relax\def\natexlab#1{#1}\fi
\expandafter\ifx\csname selectlanguage\endcsname\relax
  \def\selectlanguage#1{\relax}\fi

\bibitem[\protect\citename{Antoniak, }1974]{Antoniak}
Antoniak, C.~E. 1974.
\newblock Mixtures of {D}irichlet processes with applications to {B}ayesian
  nonparametric problems.
\newblock {\em Ann. Statist.}, {\bf 2}, 1152--1174.

\bibitem[\protect\citename{Arratia {et~al.}, }2003]{ArratiaBT}
Arratia, R., Barbour, A.~D., and Tavar\'e, S. 2003.
\newblock {\em Logarithmic Combinatorial Structures: A Probabilistic Approach}.
\newblock EMS Monogr. Math.
\newblock Zurich: European Math. Soc. Publishing House.

\bibitem[\protect\citename{Blackwell, }1973]{Blackwell}
Blackwell, D. 1973.
\newblock Discreteness of {F}erguson selections.
\newblock {\em Ann. Statist.}, {\bf 1}, 356--358.

\bibitem[\protect\citename{Blackwell and MacQueen, }1973]{BlackwellM}
Blackwell, D., and MacQueen, J.~B. 1973.
\newblock Ferguson distributions via {P}\'{o}lya urn schemes.
\newblock {\em Ann. Statist.}, {\bf 1}, 353--355.

\bibitem[\protect\citename{Donnelly and Joyce, }1989]{DonnellyJ}
Donnelly, P.~J., and Joyce, P. 1989.
\newblock Continuity and weak convergence of ranked and size-biased
  permutations on the infinite simplex.
\newblock {\em Stochastic Process. Appl.}, {\bf 31}, 89--103.

\bibitem[\protect\citename{Dunson and Park, }2007]{DunsonP}
Dunson, D.~B., and Park, J-H. 2007.
\newblock Kernel stick breaking processes.
\newblock {\em Biometrika}, {\bf 95}, 307--323.

\bibitem[\protect\citename{Escobar, }1994]{Escobar}
Escobar, M.~D. 1994.
\newblock Estimating normal means with a {D}irichlet process prior.
\newblock {\em J. Amer. Statist. Assoc.}, {\bf 89}, 268--277.

\bibitem[\protect\citename{Escobar and West, }1995]{EscobarW}
Escobar, M.~D., and West, M. 1995.
\newblock {B}ayesian density estimation and inference using mixtures.
\newblock {\em J. Amer. Statist. Assoc.}, {\bf 90}, 577--588.

\bibitem[\protect\citename{Ewens, }1972]{Ewens}
Ewens, W.~J. 1972.
\newblock The sampling theory of selectively neutral alleles.
\newblock {\em Theor. Population Biology}, {\bf 3}, 87--112.

\bibitem[\protect\citename{Ferguson, }1973]{Ferguson}
Ferguson, T.~S. 1973.
\newblock A {B}ayesian analysis of some nonparametric problems.
\newblock {\em Ann. Statist.}, {\bf 1}, 209--230.

\bibitem[\protect\citename{Finetti, }1931]{DeFinetti31}
Finetti, B.~de. 1931.
\newblock Funzione caratteristica di un fenomeno aleatorio.
\newblock {\em Atti della R. Academia Nazionale dei Lincei, ser. 6}, {\bf 4},
  251--299.
\newblock Memorie, Classe di Scienze Fisiche, Mathematiche e Naturali.

\bibitem[\protect\citename{Finetti, }1937]{DeFinetti37}
Finetti, B.~de. 1937.
\newblock La pr\'evision: ses lois logiques, ses sources subjectives.
\newblock {\em Ann. Inst. H. Poincar\'e}, {\bf 7}, 1--68.

\bibitem[\protect\citename{Gelman {et~al.}, }1995]{GelmanCSR}
Gelman, A., Carlin, J.~B., Stern, H.~S., and Rubin, D.~B. 1995.
\newblock {\em Bayesian Data Analysis}.
\newblock London: Chapman and Hall.

\bibitem[\protect\citename{Green, }1995]{Green}
Green, P.~J. 1995.
\newblock Reversible jump {M}arkov chain {M}onte {C}arlo computation and
  {B}ayesian model determination.
\newblock {\em Biometrika}, {\bf 82}, 711--732.

\bibitem[\protect\citename{Green and Richardson, }2001]{GreenR}
Green, P.~J., and Richardson, S. 2001.
\newblock Modelling heterogeneity with and without the {D}irichlet process.
\newblock {\em Scand. J. Statist.}, {\bf 28}, 355--375.

\bibitem[\protect\citename{Green {et~al.}, }2003]{GreenHSSS}
Green, P.~J., Hjort, N.~L., and Richardson, S. (eds). 2003.
\newblock {\em Highly Structured Stochastic Systems}.
\newblock Oxford: Oxford Univ. Press.

\bibitem[\protect\citename{Griffin and Steel, }2006]{GriffinS}
Griffin, J.~E., and Steel, M. F.~J. 2006.
\newblock Order-based dependent {D}irichlet processes.
\newblock {\em J. Amer. Statist. Assoc.}, {\bf 101}, 179--194.

\bibitem[\protect\citename{Hjort {et~al.}, }2010]{HjortHMW}
Hjort, N.~L., Holmes, C., M\"uller, P., and Walker, S.~G. (eds). 2010.
\newblock {\em Bayesian Nonparametrics}.
\newblock Camb. Ser. Stat. Probab. Math., vol. 28.
\newblock Cambridge: Cambridge Univ. Press.

\bibitem[\protect\citename{Holst, }2001]{Holst}
Holst, L. 2001.
\newblock {\em The {P}oisson--{D}irichlet Distribution and its Relatives
  Revisited}.
\newblock Tech. rept. Department of Mathematics, Royal Institute of Technology,
  Stockholm.

\bibitem[\protect\citename{Ishwaran and James, }2001]{IshwaranJ01}
Ishwaran, H., and James, L.~F. 2001.
\newblock Gibbs sampling methods for stick-breaking priors.
\newblock {\em J. Amer. Statist. Assoc.}, {\bf 96}, 161--173.

\bibitem[\protect\citename{Ishwaran and James, }2003]{IshwaranJ03}
Ishwaran, H., and James, L.~F. 2003.
\newblock Generalized weighted {C}hinese {R}estaurant processes for species
  sampling mixture models.
\newblock {\em Statist. Sinica}, {\bf 13}, 1211--1235.

\bibitem[\protect\citename{Ishwaran and Zarepour, }2002]{IshwaranZ}
Ishwaran, H., and Zarepour, M. 2002.
\newblock Dirichlet prior sieves in finite {N}ormal mixtures.
\newblock {\em Statist. Sinica}, {\bf 12}, 941--963.

\bibitem[\protect\citename{Kallenberg, }2005]{Kallenberg}
Kallenberg, O. 2005.
\newblock {\em Probabilistic Symmetries and Invariance Principles}.
\newblock New York: Springer-Verlag.

\bibitem[\protect\citename{Kingman, }1967]{KingmanPac}
Kingman, J. F.~C. 1967.
\newblock Completely random measures.
\newblock {\em Pacific J. Math.}, {\bf 21}, 59--78.

\bibitem[\protect\citename{Kingman, }1975]{KingmanRSS}
Kingman, J. F.~C. 1975.
\newblock Random discrete distributions (with discussion and response).
\newblock {\em J. Roy. Statist. Soc. Ser. B}, {\bf 37}, 1--22.

\bibitem[\protect\citename{Kingman, }1978]{KingmanExch}
Kingman, J. F.~C. 1978.
\newblock Uses of exchangeability.
\newblock {\em Ann. Probab.}, {\bf 6}, 183--197.

\bibitem[\protect\citename{Kingman, }1993]{KingmanPP}
Kingman, J. F.~C. 1993.
\newblock {\em Poisson Processes}.
\newblock Oxford: Oxford Univ. Press.

\bibitem[\protect\citename{Lau and Green, }2007]{LauG}
Lau, J.~W., and Green, P.~J. 2007.
\newblock {B}ayesian model-based clustering procedures.
\newblock {\em J. Comput. Graph. Statist.}, {\bf 16}, 526--558.

\bibitem[\protect\citename{Lavine, }1992]{Lavine92}
Lavine, M. 1992.
\newblock Some aspects of {P}\'olya tree distributions for statistical
  modelling.
\newblock {\em Ann. Statist.}, {\bf 20}, 1222--1235.

\bibitem[\protect\citename{Lavine, }1994]{Lavine94}
Lavine, M. 1994.
\newblock More aspects of {P}\'olya tree distributions for statistical
  modelling.
\newblock {\em Ann. Statist.}, {\bf 22}, 1161--1176.

\bibitem[\protect\citename{Lo, }1984]{Lo}
Lo, A.~Y. 1984.
\newblock On a class of {B}ayesian nonparametric estimates, {I}: {D}ensity
  estimates.
\newblock {\em Ann. Statist.}, {\bf 12}, 351--357.

\bibitem[\protect\citename{MacEachern, }1994]{MacEachern}
MacEachern, S.~N. 1994.
\newblock Estimating normal means with a conjugate style {D}irichlet process
  prior.
\newblock {\em Commun. Statist. Simulation and Computation}, {\bf 23},
  727--741.

\bibitem[\protect\citename{MacEachern, }1999]{MacEachernASA}
MacEachern, S.~N. 1999.
\newblock Dependent nonparametric processes.
\newblock {In:} {\em Proceedings of the Section on {B}ayesian Statistical
  Science}.
\newblock American Statistical Association.

\bibitem[\protect\citename{MacEachern and M\"{u}ller, }1998]{MacEachernM98}
MacEachern, S.~N., and M\"{u}ller, P. 1998.
\newblock Estimating mixture of {D}irichlet process models.
\newblock {\em J. Comput. Graph. Statist.}, {\bf 7}, 223--238.

\bibitem[\protect\citename{MacEachern {et~al.}, }2001]{MacEachernKG}
MacEachern, S.~N., Kottas, A., and Gelfand, A. 2001.
\newblock {\em Spatial Nonparametric {B}ayesian Models}.
\newblock Tech. rept. 01--10. Institute of Statistics and Decision Sciences,
  Duke University.

\bibitem[\protect\citename{McCloskey, }1965]{McCloskey}
McCloskey, J.~W. 1965.
\newblock {\em A Model for the Distribution of Species in an Environment}.
\newblock Ph.D. thesis, Michigan State University.

\bibitem[\protect\citename{M{\"u}{\SortNoop{e}}ller {et~al.}, }1996]{MuellerEW}
M{\"u}{\SortNoop{e}}ller, P., Erkanli, A., and West, M. 1996.
\newblock {B}ayesian curve fitting using multivariate normal mixtures.
\newblock {\em Biometrika}, {\bf 83}, 67--79.

\bibitem[\protect\citename{Muliere and Secchi, }2003]{Muliere}
Muliere, P., and Secchi, P. 2003.
\newblock Weak convergence of a {D}irichlet-multinomial process.
\newblock {\em Georgian Math. J.}, {\bf 10}, 319--324.

\bibitem[\protect\citename{Neal, }2000]{NealJCGS}
Neal, R.~M. 2000.
\newblock {M}arkov chain sampling methods for {D}irichlet process mixture
  models.
\newblock {\em J. Comput. Graph. Statist.}, {\bf 9}, 249--265.

\bibitem[\protect\citename{Nobile and Fearnside, }2007]{NobileF}
Nobile, A., and Fearnside, A.~T. 2007.
\newblock Bayesian finite mixtures with an unknown number of components: the
  allocation sampler.
\newblock {\em Statist. Comput.}, {\bf 17}, 147--162.

\bibitem[\protect\citename{Patil and Taillie, }1977]{PatilT}
Patil, C.~P., and Taillie, C. 1977.
\newblock Diversity as a concept and its implications for random communities.
\newblock {\em Bull. Int. Statist. Inst.}, {\bf 47}, 497--515.

\bibitem[\protect\citename{Pitman, }1995]{Pitman}
Pitman, J. 1995.
\newblock Exchangeable and partially exchangeable random partitions.
\newblock {\em Probab. Theory Related Fields}, {\bf 102}, 145--158.

\bibitem[\protect\citename{Pitman, }1996]{Pitman96}
Pitman, J. 1996.
\newblock Some developments of the {B}lackwell-{MacQueen} urn scheme.
\newblock {Pages  245--267 of:} Ferguson, T.~S., Shapley, L.~S., and MacQueen,
  J.~B. (eds), {\em Statistics, Probability and Game Theory; Papers in Honor of
  David Blackwell}.
\newblock Hayward, CA: Institute of Mathematical Statistics.

\bibitem[\protect\citename{Pitman and Yor, }1997]{PitmanY}
Pitman, J., and Yor, M. 1997.
\newblock The two-parameter {P}oisson--{D}irichlet distribution derived from a
  stable subordinator.
\newblock {\em Ann. Probab.}, {\bf 25}, 855--900.

\bibitem[\protect\citename{Richardson and Green, }1997]{RichardsonG}
Richardson, S., and Green, P.~J. 1997.
\newblock On {B}ayesian analysis of mixtures with an unknown number of
  components (with discussion and response).
\newblock {\em J. Roy. Statist. Soc. Ser. B}, {\bf 59}, 731--792.

\bibitem[\protect\citename{Sethuraman, }1994]{Sethuraman}
Sethuraman, J. 1994.
\newblock A constructive definition of {D}irichlet priors.
\newblock {\em Statist. Sinica}, {\bf 4}, 639--650.

\bibitem[\protect\citename{Sethuraman and Tiwari, }1982]{SethuramanT}
Sethuraman, J., and Tiwari, R.~C. 1982.
\newblock Convergence of {D}irichlet measures and the interpretation of their
  parameters.
\newblock {Pages  305--315 of:} Gupta, S.~S., and Berger, J.~O. (eds), {\em
  Statistical Decision Theory and Related Topics III},  vol. 2.
\newblock New York: Academic Press.

\bibitem[\protect\citename{Teh {et~al.}, }2006]{TehJBB}
Teh, Y.~W., Jordan, M.~I., Beal, M.~J., and Blei, D.~M. 2006.
\newblock Hierarchical {D}irichlet processes.
\newblock {\em J. Amer. Statist. Assoc.}, {\bf 101}, 1566--1581.

\bibitem[\protect\citename{Walker {et~al.}, }1999]{WalkerDLS}
Walker, S.~G., Damien, P., Laud, P.~W., and Smith, A. F.~M. 1999.
\newblock {B}ayesian nonparametric inference for random distributions and
  related functions (with discussion).
\newblock {\em J. Roy. Statist. Soc. Ser. B}, {\bf 61}, 485--527.

\bibitem[\protect\citename{Wen {et~al.}, }1998]{Wen}
Wen, X., Fuhrman, S., Michaels, G.~S., Carr, D.~B., Smith, S., Barker, J.~L.,
  and Somogyi, R. 1998.
\newblock Large-scale temporal gene expression mapping of central nervous
  system development.
\newblock {\em Proc. Natl. Acad. Sci. USA}, {\bf 95}, 334--339.

\bibitem[\protect\citename{West {et~al.}, }1994]{WestME}
West, M., M\"{u}ller, P., and Escobar, M.~D. 1994.
\newblock Hierarchical priors and mixture models, with application in
  regression and density estimation.
\newblock {In:} Freeman, P.~R., and Smith, A. F.~M. (eds), {\em Aspects of
  Uncertainty: A Tribute to {D}. {V}. {L}indley}.
\newblock Chichester: Wiley.

\bibitem[\protect\citename{Yeung {et~al.}, }2001]{Yeung}
Yeung, K.~Y., Haynor, D.~R., and Ruzzo, W.~L. 2001.
\newblock Validating clustering for gene expression data.
\newblock {\em Bioinformatics},  309--318.

\end{thebibliography}
\bibliographystyle{cambridgeauthordateCMG}

\end{document}